\documentclass[aps,pre,twocolumn,superscriptaddress]{revtex4}
\usepackage{amssymb}
\usepackage{bbm}
\usepackage{indentfirst}
\usepackage{graphicx}
\usepackage{amsmath,amssymb}
\usepackage{bpchem}

\usepackage{color}
\usepackage{tabularx}
\usepackage{multirow}
\usepackage{makecell}
\usepackage{array}
\usepackage{appendix}
\newcommand{\PreserveBackslash}[1]{\let\temp=\\#1\let\\=\temp}
\newcolumntype{C}[1]{>{\PreserveBackslash\centering}p{#1}}
\newcolumntype{R}[1]{>{\PreserveBackslash\raggedleft}p{#1}}
\newcolumntype{L}[1]{>{\PreserveBackslash\raggedright}p{#1}}
\begin{document}

\title{Universal scaling and phase transitions of coupled phase oscillator populations}
\author{Can Xu}\email[]{xucan@hqu.edu.cn}
\affiliation{Institute of Systems Science and College of Information Science and Engineering, Huaqiao University, Xiamen 361021, China}

\author{Xuebin Wang}
\affiliation{Institute of Systems Science and College of Information Science and Engineering, Huaqiao University, Xiamen 361021, China}

\author{Per Sebastian Skardal}
\email[]{persebastian.skardal@trincoll.edu}
\affiliation{Department of Mathematics, Trinity College, Hartford, Connecticut 06106, USA }


\newcommand{\WARN}[1]{\textcolor{green}{#1}}
\newcommand{\NOTES}[1]{\textcolor{red}{#1}}
\begin{abstract}
The Kuramoto model, which serves as a paradigm for investigating synchronization phenomenon of oscillatory system, is known to exhibit second-order, i.e., continuous, phase transitions in the macroscopic order parameter. Here, we generalize a number of classical results by presenting a general framework for capturing, analytically, the critical scaling of the order parameter at the onset of synchronization. Using a self-consistent approach and constructing a characteristic function, we identify various phase transitions toward synchrony and establish scaling relations describing the asymptotic dependence of the order parameter on coupling strength near the critical point. We find that the geometric properties of the characteristic function, which depends on the natural frequency distribution, determines the 
scaling properties of order parameter above the criticality.
\end{abstract}

\pacs{05.45.Xt, 89.75.Fb, 89.75.Hc}

\maketitle

\section{Introduction}\label{sec:01}
Synchronization of large ensembles of interacting units is a universal phenomenon that arises in a variety natural processes. Typical examples include the flashing of fireflies~\cite{ermentrout1991an}, colonies of yeast cells~\cite{richard1996ace}, pacemaker cells in the heart~\cite{taylor2010spon}, and neural activity~\cite{buzski2004sci}. Additional examples where synchronization plays a prominent role in engineered systems include power grid dynamics~\cite{filatrella2008ana} and Josephson junction arrays~\cite{benz1991coh}. Exploring the routes these systems take towards consensus and self-organization invariably begins at the onset of synchronization, and the behavior of the system thereafter remains an active area of research~\cite{kiss2002emerging}.

A remarkable prototype for studying synchronization issue is the Kuramoto model which displays such synchronization transitions~\cite{kuramoto1975}. In particular, when the coupling strength between oscillators is increased a transition from incoherence to partial synchrony, as measured by  the classical Kuramoto order parameter, takes place in a manner analogous to phase transitions observed in other physical systems~\cite{acebron2005the}. In this analogy, the phase transition between incoherence and synchronization can be characterized by a supercritical bifurcation of the order parameter. Because of its analytical tractability, the Kuramoto model together with its variant versions was extensively investigated and a great deal of progress have been made including the discovering of low-dimensional description for the order parameter~\cite{ott2008low,ott2009long}, and the identification of various coherent states on the way to synchronization\cite{abrams2004chi,abrams2008solv,xu2020bif}. However, the scaling properties of the order parameter at the onset of synchronization, i.e., at the phase transition, remains not completely understood.

In this paper, we provide a general framework for revealing the inner-relation between the phase transition and the critical behavior, i.e., scaling properties, of the order parameter in the Kuramoto model. Using a self-consistent approach, various phase transitions towards synchrony and the associated scaling behaviors of the order parameter near the onset are established in a universal form. Depending on the properties of the natural frequency distributions, we show that the bifurcation of collective dynamics involves three manners including supercritical, marginal, and subcritical that correspond to continuous (i.e., second-order), hybrid, and tiered phase transitions, respectively. In particular, we demonstrate that the structural property of characteristic function in the vicinity of its critical point determine the scaling exponents as well as asymptotic coefficients of the order parameter with leading and sub-leading terms. Our study serves as a promising strategy in unveiling the rich scaling behaviors and phase transitions in coupled oscillator networks.

The remainder of this paper is organized as follows. In Sec.~\ref{sec:02} we review the self-consistent equation for the stationary solution of the coupled phase oscillator system and introduce the characteristic function. In Sec.~\ref{sec:03} we use the characteristic function to derive a general framework for uncovering the scaling behavior of the order parameter at the onset of synchronization. In Sec.~\ref{sec:04} we give several examples that illustrate the utility of this new framework. Finally, in Sec.~\ref{sec:05} we conclude with a discussion of our results.

\section{Self-consistent approach}\label{sec:02}
We begin by considering the classical Kuramoto model in which a system of oscillators evolve according to
\begin{equation}\label{equ:01}
  \dot{\theta}_i=\omega_i+\frac{K}{N}\sum_{j=1}^N\sin(\theta_j-\theta_i), \quad i=1,\ldots,N,
\end{equation}
where $\theta_i$ is the phase of oscillator $i$, $N$ is the size of system, $\omega_i$ is the natural frequency of oscillator $i$, which is assumed to be drawn from the probability density function $g(\omega)$ with mean $\omega_0$, and $K>0$ stands for the coupling strength between oscillators. For simplicity, we may shift the mean $\omega_0$ of $g(\omega)$ to zero by entering a suitable rotating reference frame, and we will further assume that $g(\omega)$ is an even function throughout this paper. As we shall see, the symmetry assumption is essential for the bifurcation and scaling analysis below. We will also consider the continuum limit $N\to\infty$ where the population of oscillators may be described accurately by a density function.

Next, a complex order parameter (mean-field) $Z(t)$ is needed to quantified phase transition in Kuramoto model defined as
\begin{equation}\label{equ:02}
  Z(t)=R(t)\exp\{i\Psi(t)\}=\frac{1}{N}\sum_{j=1}^{N}\exp\{i\theta_j(t)\}.
\end{equation}
The amplitude $R(t)\in[0, 1]$ measures the coherence of the phases, and $\Psi(t)$ denotes the average phase of the ensemble. For the self-consistent analysis, we are interested in the steady state, which implies that $R$ is a constant and $\Psi$ rotates uniformly with $\Psi(t)=\Omega t+\Psi_0$. We however remark that a suitable global phase shift of initial conditions [under which the dynamics of Eq.~(\ref{equ:01}) are invariant] and the assumed symmetry of $g(\omega)$ makes $\Psi_0$ and $\Omega$ to be zero, respectively. As a result, the governing equation for each oscillator is simplified to the mean-field form
\begin{equation}\label{equ:03}
  \dot{\theta}_i=\omega_i-q\sin\theta_i,
\end{equation}
with $q=KR$. Depending on the natural frequencies, oscillators in Eq.~(\ref{equ:03}) can be divided into those locked by the mean-field with $\sin\theta_i=\omega_i/q$ and $\cos\theta_i=\sqrt{1-(\omega_i/q)^2}$ ($|\omega_i|\le q$) and the unlocked (drifting) ones rotating non-uniformly with period $T_i=2\pi (\omega_i^2-q^2)^{-1/2}$ ($|\omega_i|>q$). Then, the order parameter is given by
\begin{equation}\label{equ:04}
  Z=R=\langle e^{i\theta}\rangle_{lock}+\langle e^{i\theta}\rangle_{drift},
\end{equation}
where $\langle \cdot \rangle$ denotes the average over the populations. Note that, the symmetric assumptions further reduce to
\begin{equation}\label{equ:add01}
  \langle \sin\theta\rangle_{lock}=\langle e^{i\theta}\rangle_{drift}=0,
\end{equation}
whether $N$ is infinite or not. Thus, the self-consistent equation $R=\langle\cos\theta\rangle_{lock}$ in thermodynamic limit $N\to \infty$ may be written
\begin{equation}\label{equ:05}
  \frac{1}{K}=F(q)=\frac{1}{q}\int_{|\omega|<q}g(\omega)\sqrt{1-(\omega/q)^2}d\omega,
\end{equation}
where we call $F(q)$ the {\it characteristic function} of the Kuramoto model.

Eq.~(\ref{equ:05}) defines the implicit dependence of $R$ on $K$. 
When $K$ is sufficiently small the only solution to Eq.~(\ref{equ:05}) is the trivial solution $R=0$, which loses its stability at a critical point $K_a$~\cite{strogatz1991stability}. By increasing $K$, a phase transition occurs at $q_c$ corresponding to the maximum of $F(q)$ with the associated critical coupling strength
\begin{equation}\label{equ:add02}
K_c=[F(q_c)]^{-1}.
\end{equation}
Based on the structure of $F(q)$, $K_c$ is not necessarily equal to $K_a$ (we will demonstrate this later). It has been well known that for the classical KM, when $g(\omega)$ is unimodal with infinite tails, for instance Gaussian or Lorentzian distributions, a second-order phase transition takes place at $K_a=K_c=2/[\pi g(0)]$ and the bifurcating branch of $R$ near the onset obeys square-root scaling law. Our aim in the remainder of this paper is to extend this to the general case.

\section{General framework for scaling analysis}\label{sec:03}
Proceeding with our analysis, we consider a small perturbation $0<\delta q\ll 1$ of the characteristic parameter at $q_c$, or equivalently considering $K=K_c+\delta K$, $R=R_c+\delta R$ with $0<\delta K \ll 1$ and $0<\delta R\ll 1$. We then consider the expanding of $F(q)$ in powers of $\delta q$ up to the sub-leading order,
\begin{equation}\label{equ:06}
  F(q)= F(q_c)+A(\delta q)^\mu +B(\delta q)^v.
\end{equation}
Based on the property of $F(q)$ in the neighborhood of $q_c$, $\mu$ and $v$ are either integers or fractions with $0<\mu< v$ and $A$ and $B$ are the corresponding expansion coefficients. Substituting Eq.~(\ref{equ:06}) into the self-consistent equation, i.e., Eq.~(\ref{equ:05}), we have
\begin{equation}\label{equ:07}
  \delta K=-\frac{AK_c^2 \delta q^\mu +B K_c^2\delta q^v}{1+AK_c \delta q^\mu +B K_c \delta q^v}.
\end{equation}
Furthermore, expanding Eq.~(\ref{equ:07}) up to the sub-leading order of $\delta q$ leads to
\begin{equation}\label{equ:08}
  \delta K=X (\delta q)^\alpha +Y (\delta q)^\beta,
\end{equation}
with $0<\alpha<\beta$. The relation between $(A, B, \mu, v)$ and $(X, Y, \alpha, \beta)$ is determined by the specifics of the system, in particular the distribution of natural frequencies $g(\omega)$. To gain better insight into $\delta q$ in Eq.~(\ref{equ:08}), we make the following ansatz:
\begin{equation}\label{equ:09}
  \delta q=X^{-1/\alpha}(\delta K)^{1/\alpha}+H(\delta K)^\epsilon.
\end{equation}
The first term of $\delta q$ corresponds to a leading solution whereas the second term is assumed to be the perturbed solution. Inserting Eq.~(\ref{equ:09}) into Eq.~(\ref{equ:08}), we get
\begin{align}
  \delta K=&\delta K+\alpha H X^{1/\alpha}(\delta K)^{\epsilon +1-1/\alpha}+YX^{-\beta/\alpha}(\delta K)^{\beta/\alpha}\nonumber\\
  &+\beta Y HX^{1/\alpha-\beta/\alpha}(\delta K)^{\epsilon+\beta/\alpha-1/\alpha}.\label{equ:10}
\end{align}
Since $\epsilon>1/\alpha$, then $H$ and $\epsilon$ can be determined self-consistently by balancing the sub-leading term which gives
\begin{equation}\label{equ:11}
  \epsilon=\frac{\beta-\alpha+1}{\alpha},
\end{equation}
and
\begin{equation}\label{equ:12}
  H=-\frac{YX^{-(\beta+1)/\alpha}}{\alpha}.
\end{equation}
The corresponding deviation of the order parameter $R$ close to phase transition point becomes
\begin{equation}\label{equ:13}
  \delta R=\frac{\delta q}{K_c}-\frac{q_c}{K_c^2}\delta K.
\end{equation}
The asymptotic behavior of the order parameter near the onset of synchronization can then be described in the following universal form,
\begin{equation}\label{equ:14}
  \delta R=P (\delta K)^\eta+Q (\delta K)^\xi.
\end{equation}
For simplicity, we define a parameter array $\chi=(\eta, \xi, P, Q)$ for the critical exponents and asymptotic coefficients of the order parameter near $K_c$, in which $\eta$ and $\xi$ represent the leading and sub-leading scaling exponents, $P$ and $Q$ denote the corresponding asymptotic coefficients, respectively. For the case of $q_c=0$, the second term of right hand side (r.h.s.) of Eq.~(\ref{equ:13}) vanishes. Substituting Eq.~(\ref{equ:09}) into Eq.~(\ref{equ:13}) and combining it with Eq.~(\ref{equ:14}), we obtain the parameter array
\begin{equation}\label{equ:15}
  \chi=(\alpha^{-1}, \epsilon, X^{-1/\alpha}K_c^{-1}, HK_c^{-1}).
\end{equation}
Whereas $q_c>0$, Eq.~(\ref{equ:13}) shows that there exists a fixed sub-leading term for the order parameter with the scaling exponent being $1$. Hence, $\chi$ should be discussed in three distinct cases according to $\epsilon$ in Eq.~(\ref{equ:09}). For $\epsilon<1$, the second term of r.h.s. of Eq.~(\ref{equ:13}) turns out to be a higher-order term compared with Eq.~(\ref{equ:09}) and should be neglected, then $\chi$ is the same as Eq.~(\ref{equ:15}). For $\epsilon=1$, the second term of r.h.s. of Eqs.~(\ref{equ:09}) and (\ref{equ:13}) should be considered at the same time,
\begin{equation}\label{equ:16}
  \chi=(\alpha^{-1}, 1, X^{-1/\alpha}K_c^{-1},HK_c^{-1}-q_cK_c^{-2}).
\end{equation}
Likewise if $\epsilon>1$, the second term of r.h.s. of Eqs.~(\ref{equ:09}) becomes a negligible higher-order term compared with Eq.~(\ref{equ:13}), then the parameter array is determined as
\begin{equation}\label{equ:17}
  \chi=(\alpha^{-1}, 1, X^{-1/\alpha}K_c^{-1}, -q_cK_c^{-2}).
\end{equation}

\section{Examples for PTs and scaling properties}\label{sec:04}
\subsection{Continuous phase transitions}
\begin{figure}
  \centering
  \includegraphics[width=\linewidth]{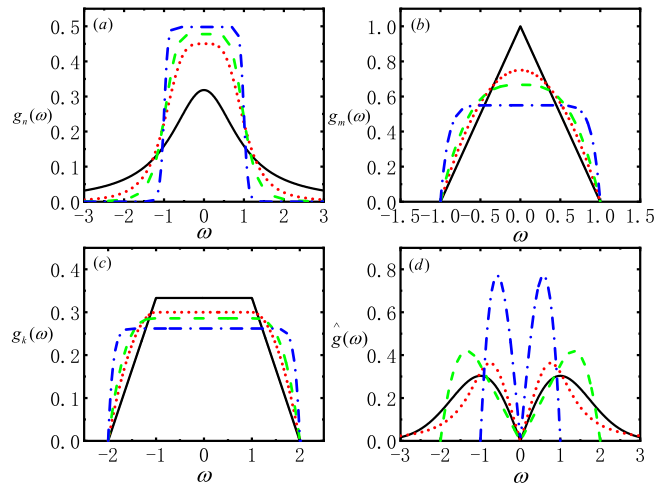}\\
  \caption{Sketch diagram of frequency distributions, (a-c) are the natural frequency distributions with $\gamma=1$ and (d) are the virtual frequency distributions. (a) $g_n(\omega)$ of Eq.(\ref{equ:18}), $n=1$ (black solid line), $n=2$ (red dot line), $n=3$ (green dash line), and $n=10$ (blue dash-dot line); (b) $g_m(\omega)$ of Eq.(\ref{equ:21}), $m=1$ (black solid line), $m=2$ (red dot line), $m=3$ (green dash line), and $m=10$ (blue dash-dot line); (c) $g_k(\omega)$ of Eq.(\ref{equ:26}) with $g(0)=C$, $k=1$ (black solid line), $k=2$ (red dot line), $k=3$ (green dash line), and $k=10$ (blue dash-dot line); (d) virtual frequency distributions of heterogenous coupling system, $\hat{g}(\omega)=|\omega|\exp\{-\omega^2/2\}/2$ (black curve line), $\hat{g}(\omega)=\frac{2|\omega|}{\pi (\omega^4 +1)}$ (red dot line), $\hat{g}(\omega)=\frac{6}{17}|\omega|[1-(|\omega|-1)^2\Theta(|\omega|-1)]$, $\omega\in [-2, 2]$ (green dash line), and $\hat{g}(\omega)=2|\omega|(1-|\omega|^2)$, $\omega\in[-1,1]$ (blue dash-dot line).}\label{fig:01}
\end{figure}

\begin{figure}
  \centering
  \includegraphics[width=\linewidth]{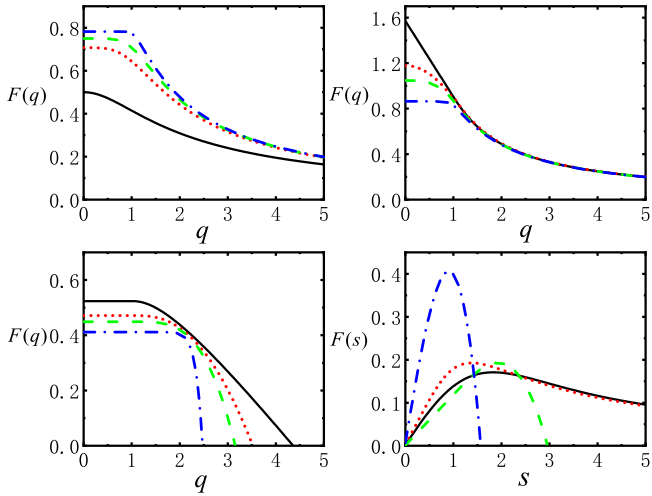}\\
  \caption{Characteristic functions are generated by the natural frequency distributions (a-c) and virtual distributions (d). (a) $F(q)$ marked with different colors corresponds to each $g_n(\omega)$ of fig.~\ref{fig:01}(a) respectively; (b) $F(q)$ marked with different colors corresponds to each $g_m(\omega)$ of fig.~\ref{fig:01}(b) respectively; (c) $F(q)$ marked with different colors corresponds to each $g_k(\omega)$ of fig.~\ref{fig:01}(c) respectively; (d) $F(s)$ marked with different colors corresponds to each $\hat{g}(\omega)$ of fig.~\ref{fig:01}(d) respectively. }\label{fig:02}
\end{figure}

As a demonstration of this approach,  we first consider $g(\omega)$ to be the following form,
\begin{equation}\label{equ:18}
  g_n(\omega)=\frac{n}{\pi}\sin\left(\frac{\pi}{2n}\right)\frac{\gamma^{2n-1}}{\omega^{2n}+\gamma^{2n}},
\end{equation}
with $\gamma>0$ and $n=1, 2,\ldots$ [see Fig.~\ref{fig:01}(a)]. As studied in \cite{skardal2018low}, $g_n(\omega)$ represents a family of rational functions in which the low-dimensional dynamics for the long-term evolution of the order parameter were obtained by means of Ott-Antonsen ansatz. The unimodality of $g_n(\omega)$ ensures that the incoherent state loses its stability via a real eigenvalue crossing the origin, which implies that $K_a=2/[\pi g_n(0)]$. Additionally, the characteristic function reduces to
\begin{equation}\label{equ:19}
  F(q)=\int_{-1}^{1} g_n(qx) \sqrt{1-x^2}dx,
\end{equation}
which is strictly decreasing in $q\in[0,\infty)$ for each $n$ [see Fig.~\ref{fig:02}(a)]. Hence, a continuous phase transition takes place at $q_c=0$ with $K_c=F(0)^{-1}=K_a$ [see Fig.~\ref{fig:03}(a)] corresponding to a trans-critical bifurcation in the parameter space.

For the scaling analysis, we note that $g_n(\omega)$ near its maximum $\omega=0$ can be expanded as
\begin{equation}\label{equ:20}
  g_n(\omega)=\frac{1}{\pi \gamma}n \sin\left(\frac{\pi}{2n}\right)\sum_{m=0}^{\infty}(-1)^m (\omega/\gamma)^{2mn},
\end{equation}
indicating that the derivative $F^{(k)}(0)\neq 0$ holds only for $k=2mn$. Power series expansion of $F(q)$ around zero leads to \begin{equation}\label{equ:add03}
  (\mu, v, A, B)=(2n, 4n, \frac{F^{(2n)}(0)}{(2n)!}, \frac{F^{(4n)}(0)}{(4n)!}),
\end{equation}
then the corresponding values of parameters in Eq.~(\ref{equ:08}) become
\begin{equation}\label{equ:add04}
  (\alpha, \beta, X, Y)=(2n, 4n, -AK_c^2, A^2 K_c^3).
\end{equation}
The order parameter $R$ near $K_c$ obeys the scaling law with the critical exponents $(\eta, \xi)=(\frac{1}{2n}, \frac{2n+1}{2n})$ and the asymptotic coefficients $(P, Q)$ are evaluated directly from Eq.~(\ref{equ:15}).

To verify its validity, we take $n=1$ as an example. In this case, $g_1(\omega)$ is a Lorentizan distribution with $K_a=K_c=2\gamma$, and the parameter array is given by
\begin{equation}\label{equ:add05}
 \chi=(\frac{1}{2}, \frac{3}{2}, \frac{4}{\sqrt{-\pi g_1^{(2)}(0)}K_c^2}, \frac{-2}{\sqrt{-\pi g_1^{(2)}(0)}K_c^3}).
\end{equation}
 Using the self-consistent equation, the exact solution of the order parameter with $K>K_c$ is $R=\sqrt{1-2\gamma/K}$. Taylor expansion of $R$ near $K_c$ yields $\delta R=(\frac{1}{2\gamma})^{1/2}(\delta K)^{1/2}-\frac{1}{2}(\frac{1}{2\gamma})^{3/2}(\delta K)^{3/2}$, which coincides with $\chi$ perfectly [see Figs.~\ref{fig:04}(a) and \ref{fig:05}(a)]. The results obtained above can be applied to arbitrary $g(\omega)$ which has integer-order of nonzero derivative at $\omega=0$ (like $g_n(\omega)\sim \exp\{-\omega^{2n}/\gamma^{2n}\}$). Therefore, we conclude that for the Kuramoto model with continuous phase transition, the asymptotic behavior of the order parameter near phase transition point is only determined by the derivable property (concave-convex) of the natural frequency distribution at its maximum.

We now extend the analysis above to a family of unimodal polynomial distributions defined in the finite support,
\begin{equation}\label{equ:21}
  g_m(\omega)=\frac{m+1}{2m}\frac{\gamma^m-|\omega|^m}{\gamma^{m+1}},\quad \omega\in [-\gamma, \gamma],
\end{equation}
with $\gamma>0$ and $m>0$ [see Fig.~\ref{fig:01}(b)]. The characteristic function $F(q)$ should be discussed in two distinct cases. If $q\leq \gamma$, we have
\begin{equation}\label{equ:22}
  F(q)=F(0)+E q^m,
\end{equation}
where $F(0)=\frac{\pi(m+1)}{4m\gamma}$ and $E=\frac{-(m+1)}{m\gamma^{m+1}}\int_0^1 x^m\sqrt{1-x^2}dx$. If $q>\gamma$, we get
\begin{equation}\label{equ:23}
  F(q)=\frac{m+1}{mq\gamma^{m+1}}\int_{0}^{\gamma}(\gamma^m-\omega^m)\sqrt{1-\omega^2/q^2}d\omega.
\end{equation}
Clearly, $F(q)$ is continuous and monotonically decreasing in $q\in[0,\infty)$ for each $m$ [see Fig.~\ref{fig:02}(b)]. Hence, the form of phase transition remains the same as above, and the system undergoes a supercritical bifurcation with exchange of the stability between the incoherence and partial synchronization at $K_a=K_c=F(0)^{-1}$.  The explicit formula of $F(q)$ makes scaling analysis easier in comparison with the series form Eq.~(\ref{equ:06}), solving the self-consistent equation yields
\begin{equation}\label{equ:24}
  R=\left(-\frac{K-K_c}{E K_c K^{m+1}}\right)^{1/m}.
\end{equation}
Expanding Eq.~(\ref{equ:24}) near $K_c$, the critical parameters for the asymptotic behavior of $R$ are $(\eta, \xi)=(1/m, 1+1/m)$, $P=(-E)^{-\frac{1}{m}}K_c^{-\frac{m+2}{m}}$, and $Q=-\frac{m+1}{m} (-E)^{-\frac{1}{m}} K_c^{-\frac{2m+2}{m}}$ [see Figs.~\ref{fig:04}(b) and \ref{fig:05}(b)].

\subsection{A hybrid phase transition}
The discussion above indicates that the phase transition corresponding to the onset of synchronization is second-order as long as there is a single maximum in the frequency distribution, and that the index of $g(\omega)$ ($n$ or $m$) controls the critical exponent of the scaling law. Increasing the index of $g(\omega)$, the continuous phase transition becomes steeper and steeper. A question here is how the order parameter in the vicinity of the critical coupling behaves as if the index tends to infinity. Because $\lim\limits_{n\rightarrow\infty}g_n(\omega)=\lim\limits_{m\rightarrow\infty}g_m(\omega)=\frac{1}{2\gamma}\Theta(\gamma-|\omega|)$ ($\Theta$ is a heaviside function), we have that $g_\infty(\omega)$ is uniform. For the purpose of scaling analysis, we further extend the limit case to a family of distributions having a plateau at the maximum, which are defined as
\begin{equation}\label{equ:26}
  g_k(\omega)=g(0)-C(|\omega|-\gamma)^k \Theta(|\omega|-\gamma)
\end{equation}
with $\gamma, \;k>0$. According to the constants $g(0)$ and $C$, the distribution is defined in a finite or infinite support with the restriction that $g_k(\omega)$ is non-negative and normalizable [see Fig.~\ref{fig:01}(c)]. As shown in refs.~\cite{basnarkov2007pha,basnarkov2008kur}, $g_k(\omega)$ represents a family of unimodal functions with a plateau section in the middle. In particular, $k=0$ and $C=g(0)$ degenerate to uniform distribution.

\begin{figure}
  \centering
  \includegraphics[width=\linewidth]{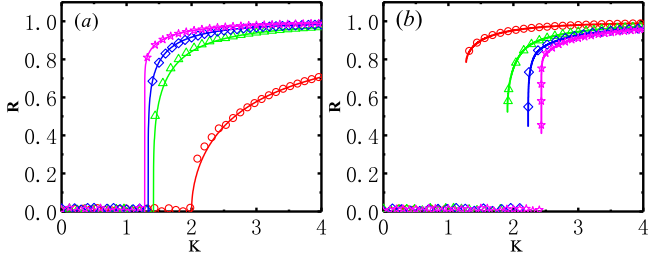}\\
  \caption{Phase diagram of the order parameter with different natural frequency distributions. (a) Continuous phase transition corresponding to $g_n(\omega)$ in Eq.~(\ref{equ:18}) with $\gamma=1$, $n=1$ (red circle), $n=2$ (green triangle), $n=3$ (blue diamond), and $n=10$ (pink star). (b) Hybrid phase transition corresponding to $g_k(\omega)$ in Eq.~(\ref{equ:26}) and the parameters are the same as fig.~\ref{fig:01}(c) with $k=0$ (red circle), $k=1$ (green triangle), $k=3$ (blue diamond), and $k=10$ (pink star). The symbols represent numerical simulations with $N=10^5$ and the solid lines correspond to the solutions of the self-consistent equation.}\label{fig:03}
\end{figure}

The characteristic function $F(q)$ corresponding to $g_k(\omega)$ should be discussed in three distinct cases, if $0\leq q\leq \gamma$, we have
\begin{equation}\label{equ:27a}
F(q)=\frac{\pi g(0)}{2};
\end{equation}
 and if $\gamma<q\leq\omega_b$,
\begin{equation}\label{equ:27b}
F(q)=\frac{2}{q}\left[\int_0^{\gamma}g(0)\sqrt{1-\frac{\omega^2}{q^2}}d\omega+\int_\gamma^q G_k(\omega, q)d\omega\right];
\end{equation}
and if $q>\omega_b$,
\begin{equation}\label{equ:27c}
F(q)=\frac{2}{q}\left[\int_0^{\gamma}g(0)\sqrt{1-\frac{\omega^2}{q^2}}d\omega+\int_\gamma^{\omega_b} G_k(\omega, q)d\omega\right].
\end{equation}
Here, we denote with $G_k(\omega,q)=[g(0)-C(\omega-\gamma)^k]\sqrt{1-\omega^2/q^2}$ and $\omega_b$ is the boundary frequency such that $g_k(\omega)=0$ for $\omega\geq \omega_b$. Remarkably, $F(q)$ is a constant for $q<\gamma$ and monotonously decreases in $q\in[\gamma, \infty)$ [see Fig.~\ref{fig:02}(c)]. Consequently, the plateau structure of $F(q)$ informs the form of the phase transition. By increasing the coupling strength, the system undergoes an abrupt transition at $q_c=\gamma$ with $K_a=K_c=\frac{2}{\pi g(0)}$, where the order parameter $R$ jumps from $0$ to a value $R_c=\frac{\pi g(0)}{2}\gamma$ [see Fig.~\ref{fig:03}(b)]. In contrast to a conventional first order phase transition~\cite{gomez2011exp}, the reversibly discontinuous transition is termed the hybrid phase transition~\cite{park2018meta}, where the hysteresis region is replaced by a vertical line at $K_c$ characterizing an infinite number of hidden metastable states formed by the oscillators with $|\omega_i|<\gamma$~\cite{mirollo2005spe}.

\begin{figure}
  \centering
  \includegraphics[width=\linewidth]{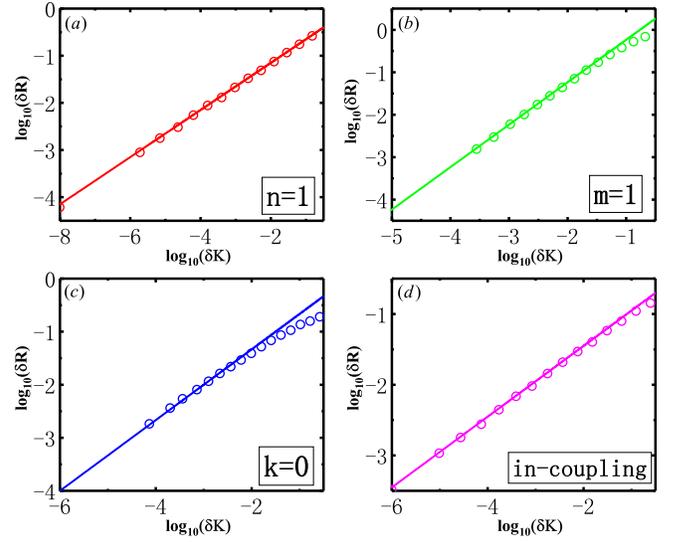}\\
  \caption{Scaling behavior of the order parameter with leading order. (a) $g_n(\omega)$ of Eq.~(\ref{equ:18}) with $n=1$ and $\gamma=1$, (b) $g_m(\omega)$ of Eq.~(\ref{equ:21}) with $m=1$ and $\gamma=1$, (c) $g_k(\omega)$ of Eq.~(\ref{equ:26}) with $k=0$ and $\gamma=0.5$, (d) in-coupling case with $\hat{g}(\omega)$ of Eq.~(\ref{equ:38}). The circles represent numerical simulations with $N=10^5$ and the solid lines correspond to analytical prediction. }\label{fig:04}
\end{figure}

\begin{figure}
  \centering
  \includegraphics[width=\linewidth]{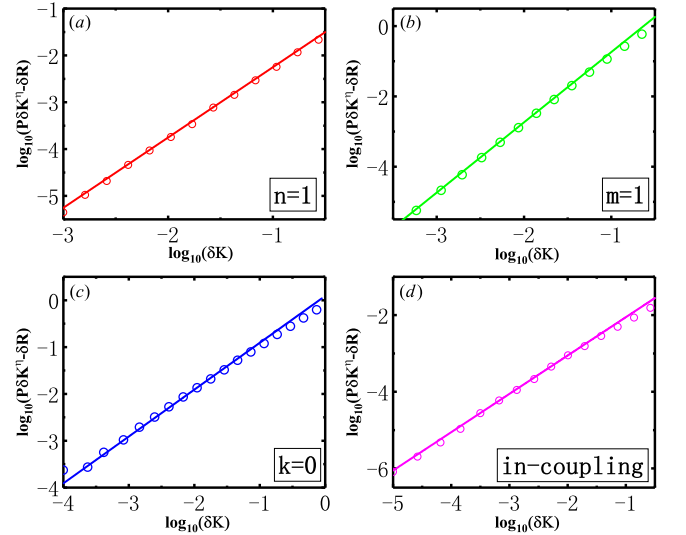}\\
  \caption{Scaling behavior of the order parameter corresponding to fig.~\ref{fig:04} with sub-leading order and the corresponding parameters are the same as fig.~\ref{fig:04}.}\label{fig:05}
\end{figure}

The property of $F(q)$ in the neighbourhood of the flat region is crucial for the scaling behavior of the order parameter. Changing variables with $\delta q=q-\gamma$ and $y=\omega-\gamma$ leads to
\begin{equation}\label{equ:28}
  F(\delta q)=\frac{\pi g(0)}{2}+\int_0^{\delta q}y^k f(\delta q,y)dy,
\end{equation}
where we have defined the function
\begin{equation}\label{equ:29}
  f(\delta q, y)=\frac{-2C}{\delta q+\gamma}\sqrt{1-\left(\frac{y+\gamma}{\delta q+\gamma}\right)^2}.
\end{equation}
After some calculations, the dominant term of deviation of $F(\delta q)$ with each order reads
\begin{equation}\label{equ:30}
\begin{split}
  F^{(n)}(\delta q)&\sim\int_0^{\delta q}y^k \frac{\partial^n f}{\partial (\delta q)^n}dy\\
  &\sim\int_0^{\delta q}y^k[1-(\frac{y+\gamma}{\delta q+\gamma})^2]^{-(n-\frac{1}{2})}dy,
\end{split}
\end{equation}
with $n=1,2,\ldots$. In the limit $\delta q\rightarrow 0$, $F^{(1)}(0)=F^{(2)}(0)=\ldots=F^{(k+1)}(0)=0$, whereas $F^{(k+2)}(\delta q)\sim (\delta q)^{-1/2}$. To avoid the divergence of the expansion coefficients, Taylor expansion of $F(\delta q)$ at $\delta q=0$ should take the following form,
\begin{equation}\label{equ:31}
  F(\delta q)=\frac{1}{K_c}+A(\delta q)^{k+\frac{3}{2}}+B(\delta q)^{k+\frac{5}{2}},
\end{equation}
and the coefficients are given by $(A, B)=(\frac{d^{2k+3}F}{dx^{2k+3}}\mid_{x=\sqrt{\delta q}}, \frac{d^{2k+5}F}{dx^{2k+5}}\mid_{x=\sqrt{\delta q}})$. Using Eq.(\ref{equ:06}), we have
\begin{equation}\label{equ:32}
  (\alpha, \beta, X, Y)=(k+\frac{3}{2}, k+\frac{5}{2}, -AK_c^2, -B K_c^2).
\end{equation}
Hence, the critical scaling exponents for the asymptotic order parameter near $K_c$ are
\begin{equation}\label{equ:33}
  (\eta, \xi)=(\frac{2}{2k+3}, \min\{1,\frac{4}{2k+3}\}).
\end{equation}
Depending on the value of $k$, the asymptotic coefficients $(P,Q)$ are determined from the standard procedure [see eqs.(\ref{equ:16}, \ref{equ:17})] .

A typical example for the illustration is to consider $k=0$ corresponding to the uniform distribution of $g(\omega)$, in which $g(0)=C=1/(2\gamma)$. The critical point for the hybrid phase transition is $(K_c, R_c)=(4\gamma/\pi, \pi/4)$, the $(\mu, v)=(\alpha, \beta)=(3/2, 5/2)$ and $(A, B)=(-\frac{2\sqrt{2}}{3}\gamma^{-5/2},
\frac{11}{5\sqrt{2}}\gamma^{-7/2} )$. The parameter array $\chi=(2/3, 1,(\frac{9\pi^7}{2^{17}\gamma^2})^{1/3}, -\frac{\pi^2}{16\gamma} )$ [see Figs.~\ref{fig:04}(c) and \ref{fig:05}(c)], which recovers the result obtained in \cite{pazo2005ther}.

\subsection{A tiered phase transition}
As we have shown above, the natural frequency distribution of Kuramoto model controls the structure of characteristic function, thereby determining the type of phase transition as well as the scaling behavior of order parameter near its onset. To better understanding this property, we generalize Kuramoto model by considering the nonuniform coupling with
\begin{equation}\label{equ:34}
  \dot{\theta}_i=\omega_i+\frac{1}{N}\sum_{j=1}^{N}K_{ij}\sin(\theta_j-\theta_i),
\end{equation}
where $K_{ij}$ is the element of the coupling matrix. In contrast to the uniform coupling in Eq.~(\ref{equ:01}), $K_{ij}$ accounts for a class of heterogeneity coupling of the oscillator system, such as the network topology~\cite{yoon2015cri}, non-local coupling~\cite{kuramoto2002}, and excitation and inhibition coupling~\cite{hong2011kur}. As demonstrated in Ref.~\cite{gao2020red}, the oscillator dynamics for certain type of heterogeneity can be reduced to Kuramoto model with uniform coupling through a remarkable transformation in the parameter space. The interplay between phase transition and the critical behavior of the order parameter can be understood in terms of a virtual frequency distribution. To this end, we adopt the coupling scheme introduced in refs.~\cite{zhang2013exp,bi2016coe,xu2018ori} establishing the frequency-coupling correlation between oscillators namely $K_{ij}=\frac{K|\omega_j|}{\langle |\omega|\rangle}$ (out-coupling) and $K_{ij}=K|\omega_i|$ (in-coupling).

Introducing the weighted order parameter
\begin{equation}\label{equ:add06}
 W e^{i\Phi}=\frac{1}{N}\sum_{j=1}^{N}\frac{|\omega_j|}{\langle |\omega|\rangle}e^{i\theta_j},
\end{equation}
the self-consistent equation of the out-coupling holds for replacing $(q, g(\omega))$ with $(s, \hat{g}(\omega))$ that are defined as $s=KW$ and $\hat{g}(\omega)=\frac{|\omega|g(\omega)}{\langle|\omega|\rangle}$. Here $\hat{g}(\omega)$ is called the virtual frequency distribution, since it plays a equivalent role of the natural frequency distribution in the Kuramoto model. This distribution combines the heterogeneity of natural frequencies and coupling, thus providing an intuitive understanding of phase transition for synchronized dynamics.  As depicted in Fig.~\ref{fig:01}(d), $\hat{g}(\omega)$ is typically bimodal for a symmetrical unimodal $g(\omega)$ with $\hat{g}(0)=\hat{g}(\pm\infty)=0$. Consequently, $F(s)$ exhibits a parabolic-like shape with $F(0)=F(\infty)=0$ [see Fig.~\ref{fig:02}(d)], and its maximum at $s_c$ with $F'(s_c)=0$.

In this setup, the bifurcation mechanism for phase transition in oscillator system becomes clear. On the one hand, $\hat{g}(0)=0$ implies that the incoherent state loses its stability via Hopf bifurcation at the critical coupling $K_a=\frac{2}{\pi \hat{g}(\Omega_c)}$~\cite{xu2016dyn,xu2019uni}, where $\Omega_c$ is the imaginary part of the eigenvalue of the linear stability analysis about $R=W=0$. On the other hand, the characteristic function $F(s)$ indicates that a saddle node bifurcation takes place at $K_c=F(s_c)^{-1}$, at which a number of oscillators lock their phases forming a macroscopic order characterized by a non-zero $W_c=s_c F(s_c)$. As the result, $K_a>K_c$ corresponds to the explosive synchronization observed previously~\cite{zhang2013exp}, whereas $K_a<K_c$ refers to the tiered phase transition where the system goes from the incoherence to the partial synchrony mediated by an oscillatory state (standing wave). The oscillatory state displaces a time-dependent coherent behavior emerging in the proximity of the critical point, where the transition from the incoherence to synchrony converts from explosive to a continuous phase transition. Such an intermediate state is born in the Hopf bifurcation way and disappears in a saddle-node infinite periodic or homoclinic bifurcation manner~\cite{martens2009exa,pazo2009exi}.

Within this framework, the power series of $F(s)$ at $s_c$ gives $(\mu, v, A, B)=(2, 3, F^{(2)}(s_c)/2,  F^{(3)}(s_c)/6)$ and the self-consistent equation leads to $(\alpha, \beta, X, Y)=(2, 3, -F^{(2)}(s_c)/[2F^2(s_c)],  -F^{(3)}(s_c)/[6F^2(s_c)])$. Thus, the asymptotic behavior of the weighted order parameter $W$ near the tiered phase transition point follows
\begin{equation}\label{equ:35}
  \chi=\left(\frac{1}{2}, 1, \frac{\sqrt{2}F^2(s_c)}{\sqrt{-F^{(2)}(s_c)}}, \frac{F^3(s_c)F^{(3)}(s_c)}{3[F^{(2)}(s_c)]^2}-s_c F^2(s_c)\right).
\end{equation}
Likewise, the order parameter $R$ near $K_c$ obeys
\begin{equation}\label{equ:36}
  R=R_c+P_R(\delta K)^{\frac{1}{2}}+Q_R (\delta K),
\end{equation}
with
\begin{equation}\label{equ:37}
  R_c=\int_{-1}^1 g(s_c x)s_c\sqrt{1-x^2} dx.
\end{equation}
The coefficients are $P_R=\sqrt{2}F(s_c)I/\sqrt{-F^{(2)}(s_c)}$ and $Q_R=F^{(3)}(s_c)F^2(s_c)I/[\sqrt{3}F^{(2)}(s_c)]^2$ with $I=\int_{-1}^1 [g(s_c x)+g^{(1)}(s_c x)x s_c]\sqrt{1-x^2}dx$.

As an analytical illustration, considering the in-coupling case, the virtual frequency distribution degenerates to a universal form
\begin{equation}\label{equ:38}
\hat{g}(\omega)=\frac{1}{2}[\delta(\omega-1)+\delta(\omega+1)]
\end{equation}
(bimodal distribution with vanishing width), which leads to a simple characteristic function $F(q)=\frac{1}{q}\sqrt{1-q^{-2}}$. The critical point corresponding to the saddle node bifurcation is $q_c=\sqrt{2}$ with $(K_c, R_c)=(2, 1/\sqrt{2})$. According to Eq.(~\ref{equ:35}), the parameter array for the order parameter $R$ near $K_c$ is
\begin{equation}\label{equ:39}
  \chi=(\frac{1}{2}, 1, \frac{\sqrt{2}}{4}, -\frac{\sqrt{2}}{16}).
\end{equation}
The analytical expression for the order parameter with $K\geq 2$ gives $R=\frac{\sqrt{2}}{2}\sqrt{1+\sqrt{1-\frac{4}{K^2}}}$ [see Figs.~\ref{fig:04}(d) and \ref{fig:05}(d)], which coincides with Eq.~(\ref{equ:39}) perfectly by imposing a power series expansion at $K=K_c$.

\section{Conclusion}\label{sec:05}
In summary, we have developed an analytical description for the critical behavior of the order parameter near the onset synchronization transition in the Kuramoto model. Based on the structures of characteristic functions in the self-consistent equation, various phase transitions were identified corresponding to different bifurcations for the emergence of collective dynamics in phase space. More importantly, this analysis provides a universal framework for uncovering the scaling properties of order parameter near the onset of synchronization. In particular, the leading and sub-leading critical exponents as well as asymptotic coefficients are all determined within this general framework. Our study provides a new way for exploring the synchronization transition in coupled oscillator systems, which could deepen the understanding of the mechanism of a phase transition in coupled dynamical networks.

\section*{ACKNOWLEDGMENTS}
This work is supported by the National Natural Science Foundation of China (Grants No. 11905068) and the Scientific Research Funds of Huaqiao University (Grant No. ZQN-810).

\end{document}